# Islamophobes are not all the same! A study of far right actors on Twitter

Bertie Vidgen, Taha Yasseri & Helen Margetts

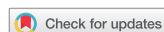

# Islamophobes are not all the same! A study of far right actors on Twitter

Bertie Vidgen [a,b], Taha Yasseri [a,b,c,d] and Helen Margetts[a,b]

[a]Oxford Internet Institute, University of Oxford, Oxford, UK; [b]The Alan Turing Institute, London, UK; [c]School of Sociology, University College Dublin, Dublin, Ireland; [d]Geary Institute for Public Policy, University College Dublin, Dublin, Ireland

**ABSTRACT**

Far-right actors are often purveyors of Islamophobic hate speech online, using social media to spread divisive and prejudiced messages which can stir up intergroup tensions and conflict. Hateful content can inflict harm on targeted victims, create a sense of fear amongst communities and stir up intergroup tensions and conflict. Accordingly, there is a pressing need to better understand at a granular level how Islamophobia manifests online and who produces it. We investigate the dynamics of Islamophobia amongst followers of a prominent UK far right political party on Twitter, the British National Party. Analysing a new data set of five million tweets, collected over a period of one year, using a machine learning classifier and latent Markov modelling, we identify seven types of Islamophobic far right actors, capturing qualitative, quantitative and temporal differences in their behaviour. Notably, we show that a small number of users are responsible for most of the Islamophobia that we observe. We then discuss the policy implications of this typology in the context of social media regulation.

## 1. Introduction

Far-right extremism risks undermining social cohesion through its divisive discourses, support for anti-democratic and violent strategies, and even physical attacks and terrorism. This can inflict harm on the vulnerable, discriminated against and/or under-represented groups targeted by the far right, such as Muslims. The Internet increasingly plays a central role within far right movements, enabling the spread, adoption and normalisation of hateful ideas and beliefs. For many far right groups, social media platforms are the lynchpin of their organisation, activism and communications. Understandably, tackling the far right has become a key policy and social concern. However, despite the scientific opportunities presented by social media data, there is a lack of large-scale research which investigates the far right online and its most harmful aspects, such as Islamophobic hate speech. In turn, policymakers, academics and civil society activists often lack the appropriate evidence to fully assess the true nature and extent of the threat posed by the far right.

We investigate the dynamics of Islamophobia amongst followers of a prominent UK far right political party on Twitter, the British National Party. We investigate the prevalence of Islamophobia and use an innovative time-scaling technique to unpick different behavioural patterns. We identify seven types of actor, capturing qualitative, qualitative and temporal differences in their behaviour. The findings offer a new way of understanding and characterising how Islamophobia manifests amongst the far right. We provide a 'meso-' view, in between broad macro-generalisations and very narrow micro-views (e.g. only analysing a small number of individuals). This research opens up a new space to evaluate far right extremism online and to theorise and tackle the social tensions and conflict which their behaviour creates.

The paper proceeds in four parts. First, we discuss existing literature on online far right behaviour and hate speech, showing that there is a lack of research into *how* Islamophobia manifests and that previous studies have not adequately considered variations in Islamophobic behaviour, despite potential policy- and theoretical benefits of doing so. Second, we describe the data and methods, including a newly collected (and publicly available) data set of 5.2 million tweets, as well as a supervised machine learning classifier and latent Markov modelling.[1] Third, we report our findings, including showing the prevalence of Islamophobia within the cohort, analysis of behaviour over time, and the identification of seven behavioural types. We then analyse the average behaviour within each type (both cross-sectionally and overtime), the number of users associated with each one, and discuss the key differences. Finally, fourth, we conclude by discussing the policy implications of our findings.

## 2. Background and research questions

### 2.1. Islamophobic hate online

Islamophobia is a deeply contested and contentious concept, and despite substantial theoretical, conceptual and critical reasoning (Bleich, 2012; Mohideen & Mohideen, 2008; Klug, 2012; Sayyid, 2014; Allen, 2013), it remains true that 'there is no singular, cogent, or consensus definition of Islamophobia' (Beydoun, 2016, p. 108). Indeed, an effort by the UK's All-Party Parliamentary Group on British Muslims to define Islamophobia as 'rooted in racism and is a type of racism that targets expressions of Muslimness or perceived Muslimness' (All Party Parliamentary Group on British Muslims, 2018) has proven highly controversial and only been partly adopted. For this paper, we use the more widely adopted academic definition of Islamophobia provided by Bleich: 'indiscriminate negative attitudes or emotions directed at Islam or Muslims' (Bleich, 2011).

Islamophobia, as with any expression of prejudice, inflicts a large human cost on its victims (Simpson, 2013; Shepherd, Harvey, Jordan, Srauy, & Miltner, 2015; Matsuda, Lawrence, Delgado, & Crenshaw, 1993; Sheridan, 2006). A 2017 review from the anti-Islamophobic charity Tell MAMA reported that 'fear of victimisation can limit the activity of British Muslims who may avoid using public transport, leaving home, or even leaving the neighbourhoods in which they feel safe' (Tell MAMA, 2017, p. 11). A report from the Muslim charity MEND also found that some victims of Islamophobia are unwilling to leave their house for fear of the abuse they may be subjected to (Ingham-Barrow, 2018). Academic research has reached similar conclusions. In a study of prejudicial abuse against Muslim communities, Allen found that although Islamophobia is often rendered 'invisible', especially when targeted against women, it can 'place substantive emotional and psychological burdens on [victims], which impinged upon their notions and experiences of feeling safe and secure' (Allen, 2015, p. 290). Other studies show the detrimmental impact of Islamophobic speech on victims' wellbeing, safety and autonomy (Awan & Zempi, 2015, 2016; Chakraborti & Zempi, 2012). It is worth noting that, even though there are important differences in the hate that takes place online, including its greater anonymity, immediacy, impact and reach (Brown, 2018), these harms can be inflicted by Islamophobia which takes place either online or offline.

The effects of Islamophobia go beyond just the immediate victims that are harmed, creating wider social impact such as undermining community cohesion and social integration and reinforcing group-based divisions. For instance, a report by the All-Party Parliamentary Group on British Muslims argued that, when left unchallenged, Islamophobia raises questions of social fairness and justice as it can have 'a significant negative impact on the life chances and quality of life enjoyed by British Muslims' (All Party Parliamentary Group on British Muslims, 2018, p. 7). Even 'casual' Islamophobia can create a breathing space for more divisive and extreme narratives to take hold.

The social implications of Islamophobia becoming normalised are much debated, with several studies highlighting the problematic associations that are routinely made between Muslims and terrorism (Allen, 2015, 2017), or between Muslims and 'barbaric' practices (Marranci, 2006; Evolvi, 2018). Such associations may be motivated by ignorance or confusion rather than prejudice, but they can contribute to a culture and status quo in which Muslims are ostracized, invisible or simply not present. Infamously, the Conservative peer Baroness Warsi alleged in 2013 that Islamophobia had passed the 'dinnertable test', with subtle forms routinely manifesting in even 'the most respectable of settings' (Runnymede Trust, 2017, v). To fully tackle the social challenges posed by prejudice, we need to tackle both structural-, legal- and economic differences between groups (e.g. a lack of legal protection or reductions in community funding and minimal access to essential services) as well as symbolic differences, such as the way in which hate crimes, including verbal abuse, are handled (Home Office, 2012, 2016, 2019).

### 2.2. Unpacking the far right

The far right has long been associated with expressing prejudice, and there is widespread consensus that one of its core features is holding exclusionary and prejudicial beliefs, which are often pursued through violent and incendiary actions, rhetoric and policies (Biggs & Knauss, 2012; Goodwin, Ford, & Cutts, 2012; Ignazi, 2003; Macklin, 2013; Veugelers & Magnan, 2005). Numerous empirical studies show the various groups that have been targeted by the far right; the activities of the British National Party alone have been variously described as racist (Richardson & Wodak, 2017), anti-Semitic (Copsey, 2007), homophobic (Severs, 2017), anti-Immigrant (Ford & Goodwin, 2010) and sexist (Gottlieb, 2017). Indeed, Mudde claims that what defines the populist right 'is natural inequality or hierarchy, not nationalism' (Mudde, 2009, p. 331), and similarly, Rydgren argues that far right parties are 'embedded in a general socio-cultural authoritarianism' (Rydgren, 2010, p. 2).

A plethora of groups have been attacked by the far right, especially with its decentralisation in recent years (Hine et al., 2017; Zannettou et al., 2017) – yet since the 9/11 attacks, the main target of their prejudice has been Muslims (Eatwell & Goodwin, 2010; Bakali, 2019; Allen, 2015). As Zúquete puts it, 'the threat that the Crescent will rise over the continent and the spectre of a Muslim Europe have become basic ideological features and themes of the European extreme right' (Zúquete, 2008). This more 'cultural' form of prejudice has in many ways replaced traditional notions of exclusionary difference within the far right, such as ideas of racism based on biologically constructed race (Rydgren, 2008). Islamophobia, having passed Warsi's 'dinnertable test', has become the acceptable public face of the far right. This framing has proven effective because Islamophobic prejudice can often overlap with genuine concerns about security, defence and integration, making it harder to identify and challenge (Falkheimer & Olsson, 2015).

The UK has long been viewed as a case of 'far right failure' (Ignazi, 2003), with the far right described as an 'ugly duckling' compared with its European counterparts (Goodwin, 2007) and stuck in 'forever a false dawn' (Goodwin, 2013). Several explanations have been put forward, including the UK's first past the post electoral system, which acts as a barrier for new parties, the lack of 'issue space' because of the ability of mainstream parties to address voters' concerns about immigration, and the lack of truly 'charismatic' far right leaders (Golder, 2016; Mudde, 2014). However, such assessments can be limited by the fact that determining the true level of support for the far right is difficult as vocalising such support can attract social condemnation. Furthermore, far right organisations are often not focused on achieving electoral success but rather on attracting attention, shaping political discourse, and attacking marginalised groups, in some cases through non-legal and violent means. Research also shows that the constituency of potential or 'latent' supporters may be far greater than the number of actual voters for the UK far right (John & Margetts, 2009) and that although small in number, far right parties have highly committed and engaged activists (Goodwin, Ford, & Cutts, 2012). Thus, even when it lacks electoral representation, the far right remains an important political and social concern.

The most prominent far right party in the UK's history is the British National Party (the BNP), which was founded in the early 1980s. During the 2000s, several BNP councillors and a London assembly member were elected, as well as two Members of the European Parliament in 2009. At the 2010 general election, the BNP 'broke through' and received half a million votes, which was 1.9% of the total (although no BNP Members of Parliament were elected). However, since then the BNP has suffered setbacks as the far right landscape has diversified and the party has faced internal troubles, bankruptcy and legal proceedings (Hope Not Hate, 2017). Despite its poor performance in recent elections, the BNP has long been a hallmark of the far right landscape and shares ideological similarities with newer and less organisationally stable far right groups (such as Generation Identity and the 'Casuals'). Even though it is unlikely that far right groups in the UK will achieve electoral success in the near future, their impact should not be underestimated given their considerable reach on social media and across society, and their ability to use this to cause social division and conflict, especially by sharing hateful messages (Home Office, 2018; Commission for Countering Extremism, 2019).

### 2.3. Understanding the dynamics of online Islamophobia within the far right

There is consensus that Islamophobia is a core part of far right ideology and online discourse (Evolvi, 2018; Froio, 2018). However, less attention has been paid to how it manifests and whether there is heterogeneity *within* the far right. This is a key omission given that such insights would help us to better characterise the far right, understand the true nature and extent of the threat that it poses, identify people who are vulnerable to far right radicalisation, develop better support for victims and also create more nuanced theoretical explanations. In particular, such research could assist with developing targeted interventions to reduce intergroup violence and minimise conflict, especially in times of heightened social tension (Leader Maynard & Benesch, 2016).

Existing research provides competing accounts of whether, and how, the far right exhibits internal heterogeneity, especially in relation to hate and prejudice. Much of the traditional 'offline' literature suggests that the far right comprises a homogenous block of like-minded and prejudiced individuals. For instance, Trilling describes supporters of the BNP as 'bloody nasty people' (Trilling, 2012) and Biggs and Knauss use membership of the BNP as a proxy for holding prejudicial beliefs (Biggs & Knauss, 2012). Similarly, in a review of research into the far right, Rydgren describes the 'ethnic competition thesis' as a key explanation of voting for far right parties because 'even if not all voters who hold anti-immigration attitudes vote for a new radical right-wing party, most voters who do vote for such parties hold such attitudes' (Rydgren, 2007, p. 250). This position is supported by Golder in a subsequent review, who discusses how ethnic competition drives far right support through economic and cultural grievances (Golder, 2016, pp. 483–485). These arguments have some support within online-specific research. In a measurement study, Chandrasekharan et al. suggest that all of the content posted on Reddit, the social news aggregation website, by members of certain banned subreddits, including r/fatpeoplehate and r/CoonTown, is toxic (2017). Awan also describes how far right actors use social media 'to inflame religious and racial tensions' by creating 'walls of hate' (Awan, 2016). These accounts suggest that most far right actors can homogenously be characterised as deeply and vocally prejudiced individuals. However, this risks characterising the entire far right on the basis of a vocal minority of actors and potentially ignores key differences they exhibit.

At the same time, other studies suggest that the far right is far more internally heterogeneous. For instance, far right voters can be motivated by economic deprivation and a desire to 'protest' against mainstream parties rather than just prejudice (Cutts, Ford, & Goodwin, 2011). Far-right supporters also sometimes express ambiguous and non-prejudicial views towards outgroups (Rhodes, 2011). Morrow and Meadowcroft investigate the organisation of the English Defence League through its rise and fall, and find that as well as having deeply committed members, it was primarily comprised of numerous 'marginal members' who had 'low levels of commitment' (Morrow & Meadowcroft, 2019). Research from the field of Internet studies suggests that online hate varies substantially across users, time, context and geography (Bliuc, Faulkner, Jakubowicz, & McGarty, 2018; Chatzakou et al., 2017; Hine et al., 2017). In a study of Islamophobic tweets sent during 2016/2017, researchers at Demos found that of all the hateful tweets they collected ∼15% were sent by 1% of the most active users and 50% were sent by just 6% of the most active users (Demos, 2017). This means that, although there is a core of highly active users, many people who engage in Islamophobic hate do so infrequently – and as such may not be deeply committed or ideologically prejudiced. These important differences in the volume of Islamophobic hate created by far right individuals are rarely considered explicitly when the far right is being assessed. Similarly, Burnap and Williams show that online hate follows temporal dynamics, exhibiting peaks and troughs around contentious 'trigger' events, such as terrorist attacks (Williams & Burnap, 2016).

From a more qualitative perspective, Awan and Zempi also write that the far right 'exploit the virtual environment and world-wide events to incite hatred towards Islam and Muslims' (Awan & Zempi, 2017). These findings suggest that producing online hate is not a stable feature of individuals which simply reflects some underling prejudice but, instead, a behaviour that can be activated or suppressed over time. Indeed, Ganesh, reflecting on the fluctuating, dispersed and complex nature of online hate and the far right, argues that it should be conceptualised as a 'swarm' with constantly shifting allegiances, priorities and levels of commitment – rather than a single goal of spreading hate (Ganesh, 2018). Similarly, Baele et al. present an analytical framework for analysing the far right online and summarise that it comprises a 'dynamic and rapidly evolving ecosystem', noting its size, breadth and heterogeneity (Baele, Brace, & Coan, 2020). These arguments also reflect broader arguments made in online research that online spaces are decentralised, complex and disruptive, and as such are far more uncertain than offline spaces (Margetts, John, Hale, & Yasseri, 2015).

The degree to which the far right exhibits variation in how Islamophobia is expressed, particularly online, remains an open question, with our understanding of this issue limited by a lack of quantified data-driven research. To address this research gap, we investigate the dynamics of Islamophobia amongst followers of the BNP on Twitter over a period of one year, using a data set of 5.2 million tweets. We are guided by a single RQ which will advance scholarship in this area:

> RQ: How does Islamophobic behaviour manifest amongst far right actors on Twitter?

## 3. Data and methods

### 3.1. Data

All tweets sent by followers of the BNP's Twitter account (@bnp) were collected from 1 April 2017 to 1 April 2018. This period covers several important political events in the UK, including the General Election on 8 June 2017, Local Elections on 4 May 2017, Manchester Arena bombing on 22 May 2017, London Bridge terror attack on 3 June 2017 and the progression of the European Union (Withdrawal) Act of 2018 through the UK parliament. Tweets were collected using Twitter's Search API, which allows a maximum of 3200 tweets to be collected from each user's timeline (including retweets). We collect data on a weekly basis and only miss tweets from users who exceed this high weekly limit.

At the start of the period (1 April 2017), there were 13,002 followers of the BNP, and at the end (31 March 2018), there were 13,951. Of the original 13,002 users, 11,785 (90.6%) were still followers at the end (1,217 ceased following). Given that it is easy to start and stop following accounts on social media, often indicating a lack of genuine interest, we only include users in the data set who follow the BNP across the entire period; 5310 of these 11,785 users (45%) tweeted at least once during the period. We remove tweets which were sent in languages other than 'English' or 'Undetermined'. This reduces the number of users to 5242.

We remove users who exhibit bot-like behaviour, which we define as accounts who send more than 40 tweets per day on average. This approach can be understood as a way of removing high-activity users, including both bots and genuine users with idiosyncratic or semi-automated tweeting patterns (Larsson & Hallvard, 2015). About 114 users meet this bot-like detection criterion and are removed, leaving 5128 users in the data set.[2] The 114 users differ substantially from the other 5128 users in several ways. First, on average they send 23,241 tweets during the period, compared with 1018 for the other users. Second, the accounts tend to be older, typically having been created 209 days earlier. Third, none of the 114 bot-like users provides their geographic location in their profile whilst 87 of the remaining 5128 do so (1.7%). These differences suggest that our bot removal method has identified a subset of distinct and highly atypical users. We verify this by applying the 'botometer' from researchers at the University of Indiana, which uses 1000 features in a supervised learning algorithm to calculate the probability that users are bots (Varol, Ferrara, Davis, Menczer, & Flammini, 2017). The 114 users we identify as exhibiting bot-like behaviour have an average bot probability of 0.630 whereas the remaining 5128 have a probability of 0.289. After all removals, the final data set consists of 5,221,256 tweets.

### 3.2. Measurement of Islamophobia

We use a machine learning classifier to detect Islamophobia which is based on Bleich's definition (provided above). It assigns tweets to one of three classes on an ordinal scale: None, Implicit and Explicit Islamophobia (Vidgen & Yasseri, 2020). The Explicit/Implicit distinction is widely used in research detecting abusive content online (Vidgen et al., 2019) and enables granular insight into the different ways in which Muslims, and Islam, are prejudicially associated with negative traits (Benford & Snow, 2000). The classifier primarily uses a word embeddings model, GloVe, as well as other relevant textual and non-textual input features, such as Part of Speech tags, to identify different types of Islamophobic content. On a held-out test data set, it has a balanced accuracy of 0.83 and a micro-F1 score of 0.78. Precision, recall and accuracy are above 0.7, which indicates it can be applied empirically (van Rijsbergen, 1979). More detail is provided in the Methods Appendix.

### 3.3. Latent Markov modelling

Latent Markov (LM) modelling is an extension to the traditional Markov chain model which can be used to model the temporal dynamics of individual behaviour and as such is suitable for addressing our Research Question. It assumes the existence of *K* latent states, where *K* must be defined in advance (Spedicato & Signorelli, 2013). The LM model then estimates both the behaviours associated with each latent state, the transitional probabilities between states, and the state which each user is assigned in each time period. Parameters are estimated using maximum likelihood estimation via the expectation–maximization algorithm (Bartolucci, Farcomeni, & Pennoni, 2010). We fit our model with time-homogeneous transitional probabilities (i.e. the transition probabilities are constant over all time periods).

Studying users' behaviour on Twitter longitudinally is difficult because users tweet at irregular times. Therefore, the actual timestamps of tweets cannot be used as this would create an LM model with millions of different 'events', a few of which line up with each other. One solution is to measure tweets within a pre-defined time window, such as 1 d. However, this risks introducing considerable biases because users send different volumes of tweets over time. As such, we scale the time by the frequency of tweets. The overall timeline of 5,221,256 tweets is divided into 100 periods, each of which consists of 52,213 tweets. The amount of linear time that each time period covers range from 1.7 days to 8.7 days. This approach is counter-intuitive but ensures that (i) for each user, the number of time periods without a value is minimised and (ii) users are compared across the same time intervals; $t_x$ covers the same time period for every user – it is just that the linear length of $t_x$ is not the same as the linear length of $t_{x+1}$. Given that the choice of 100 periods is arbitrary, we run our models with 10, 25, and 50 time periods and report similar results. Details of fitting for the LM model are given in Methods Appendix.

Using the classifier introduced by Vidgen and Yasseri (2020), we classify each tweet into one of the three levels of an ordinal variable: None, Implicit, and Explicit Islamophobia. For the LM model, we measure Islamophobia for each user in each time period by taking just the highest class of tweeting they exhibit. For instance, if a user sends at least one tweet that is Explicit Islamophobic during $t_x$, then that is how their behaviour is characterised in $t_x$. If they send at least one Implicit Islamophobic tweet but no Explicit tweets, then their behaviour is characterised as Implicit. It is only characterised as None if they send no Implicit or Explicit tweets. This strategy ensures that Islamophobic tweets are well represented in the LM model. It is also theoretically robust since what is of greatest interest is whether users have engaged in Islamophobic behaviour rather than whether, for instance, the *majority* of their behaviour is Islamophobic. We check this method for measuring Islamophobia by comparing it with a model where we only assign each user in each time window to the Islamophobic classes if at least 5% of the tweets are Islamophobic. This produces very similar results, primarily because most users send low volumes of tweets in each time window, which means that if there is at least one Islamophobic tweet, then it usually accounts for more than 5%. Even though we use a varying time period scaled by the overall volume of tweets, some users do not send any tweets in some time periods. Rather than treating these periods as missing, we assign them a value of None Islamophobic.

## 4. Results

### 4.1. Prevalence of Islamophobia

Of the 5.2 million tweets in the data set, 4.4 million are non-Islamophobic (83.8%), 0.57 million are Implicitly Islamophobic (10.8%) and 0.28 million are Explicitly Islamophobic (5.3%). Twice as much of the Islamophobia expressed by followers of the BNP is Implicit (i.e. subtle and nuanced) rather than Explicit (i.e. aggressive and overt). This provides evidence that even amongst far right actors, who are on the extremes of political discourse, Islamophobia is more likely to be implicit and covert rather than explicit. This is shown in Figure 1(A).

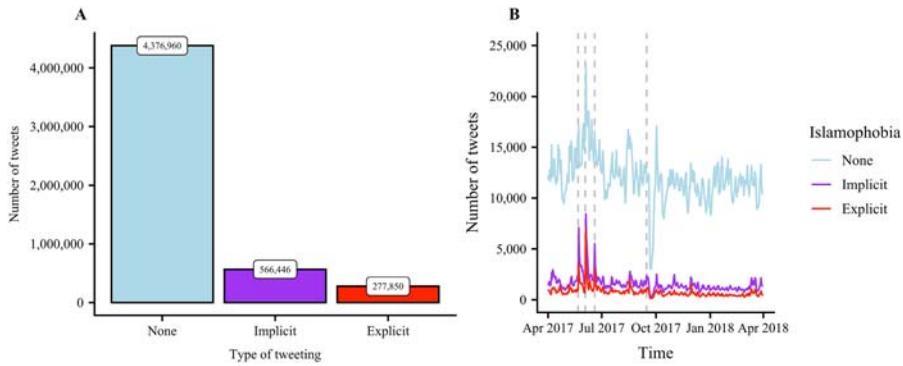

**Figure 1.** (A) Prevalence of Islamophobic tweets within 5.2 million tweet data set, (B) Prevalence of Islamophobic tweets overtime for all users in the cohort. Grey dotted lines show terror attacks in the UK.

The prevalence of Islamophobia fluctuates considerably during the period studied. There are several peaks, most noticeably at the start when several terror attacks took place. Previous research indicates these are likely to drive spikes in online hate and may explain some of the variations observed here (Williams & Burnap, 2016). This lends further support to our decision, described in Methods, to split the data into 100 'time' windows of 52,213 tweets rather than using linear time. This approach accounts for the huge variations in the total number of tweets sent during the period studied.

The prevalence of Islamophobia over time across the whole cohort of users is shown in Figure 1(B). The grey dotted lines indicate UK terror attacks in Manchester, London Bridge, Finsbury Park, and Parsons Green on 22 May, 3 June, 19 June, and 15 September 2017. Note that on 8 June 2017, a general election was held.

The distribution of tweets per user is long-tailed, as shown in Figure 2(A). The maximum number of tweets is curtailed at 14,600 because during the sampling process high activity users are removed. The distribution of Islamophobic tweets per user (combining Implicit and Explicit) is shown in Figure 2(B) and is also long-tailed.

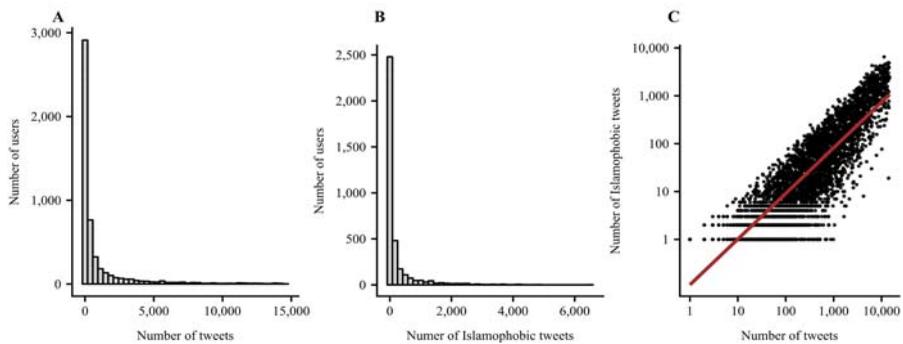

**Figure 2.** (A) Distribution of all tweets for all users in the cohort, (B) distribution of Islamophobic tweets (implicit and explicit combined) for all users in cohort, and (C) scatter plot of the number of Islamophobic tweets against the total number of tweets per user. In (C) axes are logarithmic and users who do not send any Islamophobic tweets ($n = 1484$) are not shown.

The Gini coefficients for the distribution of Implicit Islamophobia among users are 0.812, Explicit Islamophobia is 0.803, and both combined is 0.822. The Gini coefficient for the distribution of all tweets is very similar, 0.806. Overall, a small number of users are responsible for most of the Islamophobic tweets in the data set. Figure 2(C) shows the number of Islamophobic tweets versus the total number of tweets sent by each user.

These analyses suggest that the overall prevalence of Islamophobia reported in Figure 1(A) offers only a very coarse characterisation of users' behaviour. Figure 1(B) shows that there are considerable variations over time and Figure 2 that there are variations in terms of how much Islamophobia each user sends, both as an absolute value and as a proportion of the total number of tweets sent (see Figure 2(C)). Taken together, these findings suggest that patterns of Islamophobic behaviour are highly heterogeneous and that time is likely to be an important aspect for understanding how users differ. For instance, some users may follow the overall temporal trend shown in Figure 2(B) whilst others will diverge from it. These variations are investigated further in the next section.

### 4.2. Islamophobic types: disagreggating users in the cohort

To better understand the differences between users, we identify distinct user types from the data. A 'user type' can be understood as a typified dynamic pattern of behaviour followed by a subset of users in the cohort over time. It is akin to a pathway, as has been examined in studies of terrorism (McCauley & Moskalenko, 2008), and a customer journey, as studied in business and management research (Marquez, Downey, & Clement, 2015). To model the user types, we fit an LM model with $K = 3$ over the 100 time periods (shown in Figure 3).

Based on initial exploratory analyses, we separate two groups of users before fitting this model: those who send only None Islamophobic tweets ($n = 1484$) and those who send only None and Implicit Islamophobic tweets ($n = 718$). The LM model is fit on the remaining users ($n = 2,926$), all of whom send tweets across the three classes of None, Implicit, and Explicit Islamophobia.

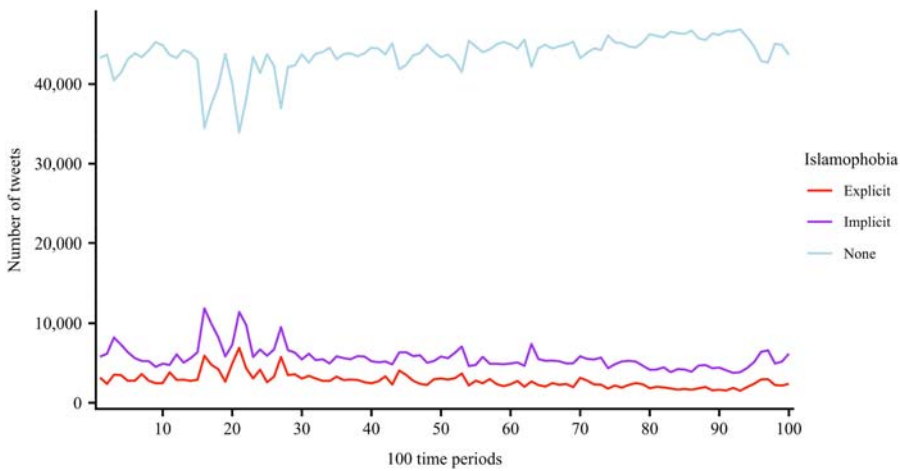

**Figure 3.** Time series for cohort of users, data represented with 100 fixed intervals of 52,213 tweets.

**Table 1.** Behavioural probabilities for latent states.

| Islamophobia | State 1 | State 2 | State 3 |
| --- | --- | --- | --- |
| None | 0.95 | 0.45 | 0.06 |
| Implicit | 0.03 | 0.32 | 0.13 |
| Explicit | 0.02 | 0.22 | 0.81 |
| Total | 1 | 1 | 1 |

The three latent states in the LM model reflect different propensities to engage in each type of tweeting. State 1 has a 0.95 probability of None Islamophobic and low probabilities for both Implicit and Explicit Islamophobia (0.03 and 0.02): when users are in this latent state they are overwhelmingly likely to not engage in any Islamophobia. State 2 is the most evenly distributed across the three types, with probabilities that range from 0.22 to –0.45: users in this latent state will exhibit highly varied behaviour. State 3 has a 0.81 probability for Explicit Islamophobia, 0.13 for Implicit and just 0.06 for None: users in this latent state are highly likely to send an Islamophobic tweet, particularly an Explicit one. These probabilities are shown in Table 1.

Each state has a transitional probability, which captures how likely users are to either stay in the same state or move states. These are shown in Table 2. The probabilities are all very high for staying in the same state (0.95 to –0.99). However, interestingly, users in States 2 and 3 are more likely to shift to a different state than users in State 1. There is a probability of 0.95 and 0.96 of staying in States 2 and 3, respectively, whilst there is a probability of 0.99 of staying in State 1. This suggests that Islamophobic tweeting is a less stable behaviour compared with None Islamophobic tweeting.

The LM model provides a simplified representation of the underlying data, in which each user is represented as a vector of length 100 (in line with the 100 time periods), each value of which is a latent state. We cluster these vectors using the *k*-modes clustering algorithm. Through fitting to minimise within the sum of squares, and manual inspection, we identify that five clusters are optimal (see Methods Appendix). Each cluster represents a certain trajectory among the three latent states, over time. Each user is assigned to one of these clusters and cannot move between them. We call each cluster a user type. In addition, there are two further user types, which comprise the two groups separated at the start: (1) users who send only None Islamophobic tweets and users who send only None and Implicit Islamophobic tweets. In total, we identify seven user types within the entire cohort.

To analyse the differences between user types, we calculate a metric called Type Score (*TScore*). For each user type, in each time period, we take the average number of tweets for each type of tweeting (None, Implicit and Explicit Islamophobia) and divide by the average number of tweets (in each type of tweeting) across the whole cohort. This is shown in Equation (1), where $p$ denotes the time period, $I$ is the type of tweeting, $U$ is the number of users, *u* denotes the user, *T* is the user type, and *n* is the number of tweets.

**Table 2.** Transitional probabilities for latent states.

|  |  | State at $t_{i+1}$ | | |
| --- | --- | --- | --- | --- |
|  |  | 1 | 2 | 3 |
| State at $t_i$ | 1 | 0.99 | 0.01 | 0.00 |
|  | 2 | 0.03 | 0.96 | 0.01 |
|  | 3 | 0.02 | 0.03 | 0.95 |

$$\mu_{pl} = \frac{\sum_{u=1}^{U} n_{upl}}{U}$$

$$T_{pl} = \frac{\sum_{u=1}^{U_T} n_{upl}}{U_T}$$

$$TScore_{pl} = \left(\frac{T_{pl}}{\mu_{pl}}\right) \times 100$$

TScore can be interpreted as a coefficient where 100 indicates the user type is in line with the average of the whole cohort and any other value is a multiple of the cohort average, indicating a deviation from this overall trend. For instance, a value of 25 for Implicit Islamophobia indicates that users in this type send 25% of the average amount of Implicitly Islamophobic tweets across the whole cohort. A value of 200 indicates that 200% of the average has been sent.

The behavioural patterns of the seven types are shown in Figure 4. In each panel, the average prevalence of tweeting across the whole cohort (for each time period and within each type of tweeting) is depicted by the horizontal grey dashed line, which is at a constant of 100. We name the seven types: None, Very Low, Low, High, Very High, Escalating and De-escalating. The seven types capture differences in the volume of tweets which users send, the strength of those tweets, and their temporality, and are described in Table 3. Noticeably, we do not identify a Moderate types as none of them have TScores which are consistently close to 100.

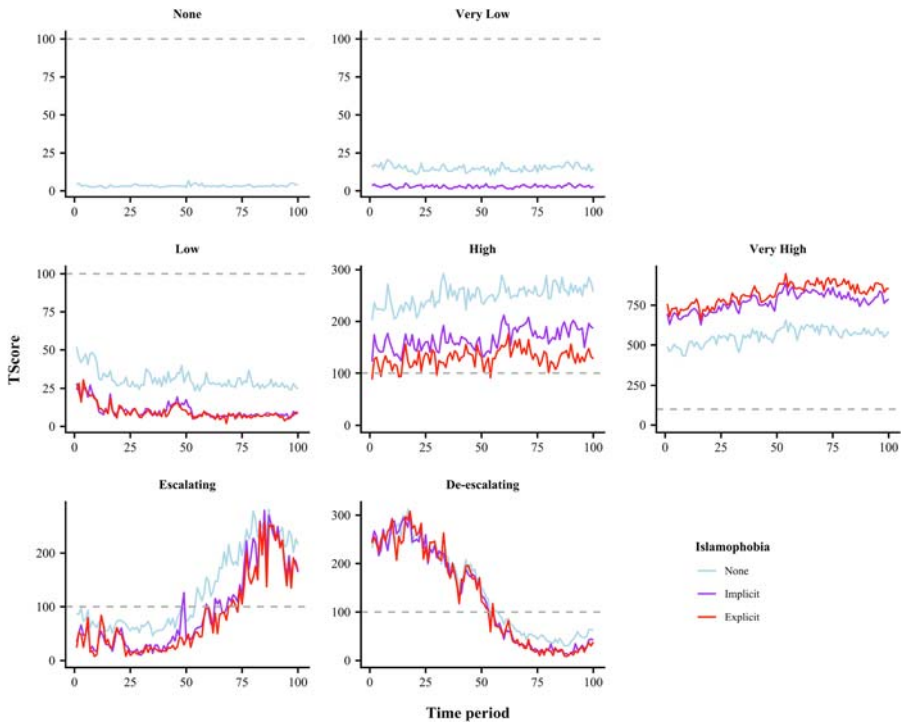

**Figure 4.** TScores for the seven types of Islamophobia over the 100 time periods. The grey dotted line shows the average tweeting for the entire cohort. Vertical axes' scales are free to vary.

**Table 3.** Descriptions of the seven users types.

| Name | Description |
| --- | --- |
| None | Users who never engage in any form of Islamophobia (whether Implicit or Explicit) |
| Very Low | Users who engage in very little Implicit Islamophobia and no Explicit Islamophobia |
| Low | Users who engage in both Implicit and Explicit Islamophobia, far below the average level |
| High | Users who consistently engage in an above-average level of Implicit and Explicit Islamophobia |
| Very High | Users who consistently engage in a high level of Islamophobia and are comparatively more likely to engage in Explicit |
| Escalating | Users whose activity, including the proportion of Islamophobia, is increasing over time |
| De-escalating | Users whose activity, including the proportion of Islamophobia, is decreasing over time |

The TScores for five of the seven types (None, Very Low, Low, High and Very High) are broadly stable over time, and the changes in magnitude are only small (i.e. for Low the TScores decrease slightly over time and for Very High they slightly increase). These types primarily capture differences in the volume of tweeting, as well as some qualitative differences. For instance, the Very High type is the only type in which the TScores for Implicit and Explicit Islamophobia are greater than the TScore for None. This indicates that users in this type engage in more Islamophobic behaviour both in absolute terms and as a proportion of their total behaviour. Similarly, the relationship between the Islamophobic classes of tweeting (Implicit and Explicit Islamophobia) differs across types. For Low, the TScores are closely aligned, for High, the TScore for Implicit is greater than for Explicit, and for Very High the TScore for Explicit is greater than Implicit. The consistency of users' behaviour also varies between types. The High type has greater variance over time than the Low and Very High types. This suggests these user types are more susceptible to exogenous shocks, which could be driving short-term fluctuations.

The Escalating and De-escalating types differ from the other types because the users of these types show a clear change in behaviour. For both Implicit and Explicit Islamophobia, the TScores for the Escalating type start from below average (∼30) and finish far above (∼170). In contrast, in the De-escalating type, the TScores for Implicit and Explicit tweeting are high at the start (∼240) and far lower at the end (∼40). In Figure 4, the time series for both Islamopohobic behaviours in the Escalating and De-escalating types appear to largely follow the time series for None Islamophobic. To better uncover this, we plot the differenced TScore values, shown in Figure 5, which shows more clearly the qualitative change in behaviour observed between the users in each type over time; for the Escalating type, the Islamophobic behaviours (both Implicit and Explicit) increase relative to the None Islamophobic behaviour whilst for De-escalating, the opposite is observed. Overall, the seven user types capture substantial quantitative, qualitative, and temporal differences between users.

Quantitative differences between the types are shown in Figure 6, which depicts the mean and dispersion of the number of each type of tweets for each type. The statistics are provided in Methods Appendix. Noticeably, the Escalating and De-escalating types have similar average levels of tweeting in each class (even though their temporal pathways are very different) showing the importance of taking time into account explicitly. Overall, the differences between the types are highly statistically significant. We conduct non-parametric omnibus tests of statistical significance: ANOVA type, Wilks' Lambda type, Lawley Hotelling type and Bartlett Nanda Pillai type, as well as permutation variations (Ellis, Burchett, Harrar, & Bathke, 2017). These are all statistically significant ($p < .000001$). We further verify this with Kruskall–Wallis tests on each of the three types of tweeting. In all cases, the differences are statistically significant ($p < .000001$). We then conduct pairwise Wilcoxon rank sum tests on each pair of types, and all differences are significant ($p < .000001$). These results provide strong evidence that differences between types are significant.

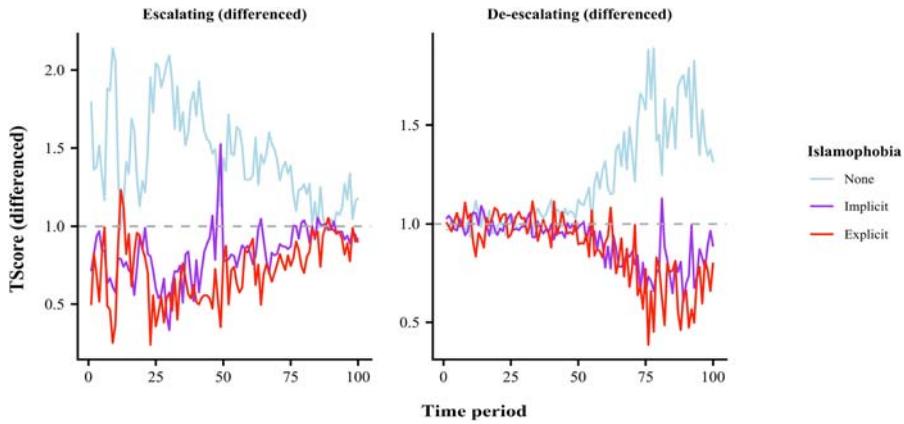

**Figure 5.** Differenced TScores for Escalating and De-escalating user types over time.

The prevalence of each user type varies considerably, as shown in Figure 7. The most prevalent is None, which accounts for 28.9% of users. Very Low comprises 14.0%. It is plausible that many of the users in these types (total, 42.9%) are less committed to far right politics. They may be followers of other political parties, journalists or academics. The Low type comprises a further 27.0% of users. In contrast, the 18% of users in the

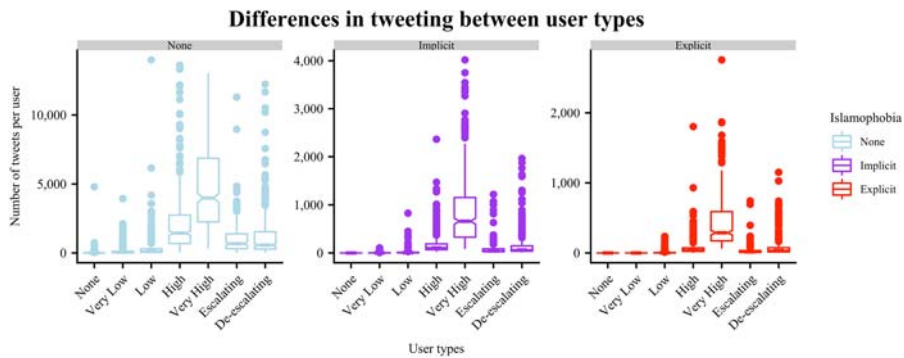

**Figure 6.** The mean and dispersion of each type of tweeting (None, Implicit and Explicit) for the seven types of Islamophobia.

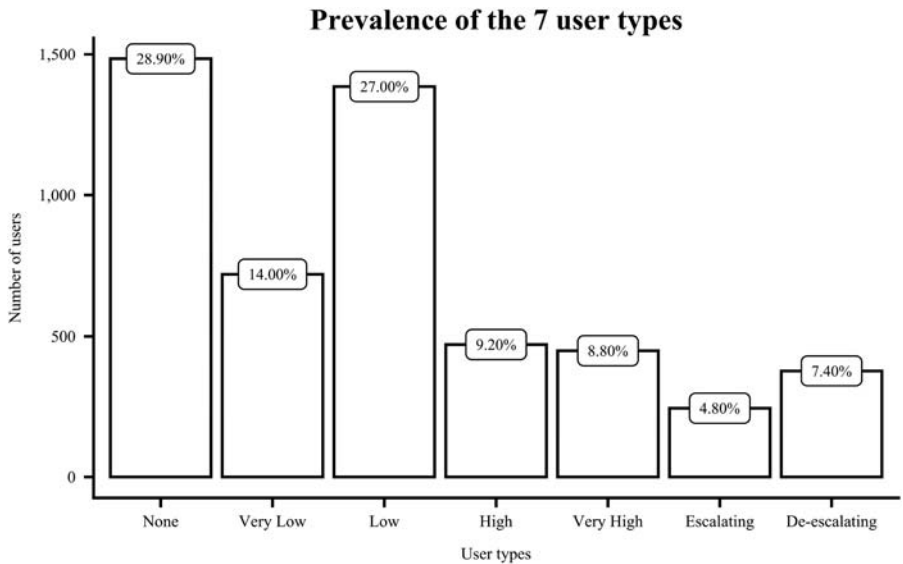

**Figure 7.** The number and percentage of users assigned to each of the seven user types.

High and Very High types (9.2% and 8.8%, respectively) are perpetually engaging in considerable levels of Islamophobia. Finally, the two types with the most noticeable temporal trends, Escalating and De-escalating, comprise 4.8% and 7.4%, respectively. There are 50% more De-escalating compared to Escalating users. The greater proportion of De-escalating most likely reflects the fact that several Islamist terrorist attacks occurred at the start of the period, which might have motivated some users to send many Islamophobic tweets in response (Burnap et al., 2014).

## 5. Discussion and conclusion

This paper has addressed a single research question ('How does Islamophobic behaviour manifest amongst far right actors on Twitter?'), providing a data-driven window into the hetereogenous and complex dynamics of Islamophobic tweeting. We have identified seven distinct user types, which capture quantitative, qualitative and temporal differences in users' behaviour. This effectively provides a 'meso' view on the far right, situated in between an overly broad 'macro' view (i.e. where an average is used to summarize a phantom 'typical' user) and a super-close up 'micro' view (i.e. where each individual is analysed separately). This lens opens up a new space in which to evaluate the far right, consider how different actors within it contribute to social tensions and, ultimately, identify ways in which these tensions can be ameliorated. These results can be used to inform future areas of research, such as investigating how the motivations and goals of people in each of the seven types differ.

Our findings support the view that the followers of the BNP on Twitter comprise a shifting, complex assemblage of individuals, as proposed by Ganesh's concept of the 'swarm' (Ganesh, 2018). They also resonate with the argument presented by Morrow and Meadowcroft apropos the organisational composition of the far right, and its historical reliance on many 'marginal members' who are not all deeply invested in far right ideology (Morrow & Meadowcroft, 2019). We show that (a) users' behaviour is captured poorly by summary statistics; Islamophobia varies across time, often exhibiting sharp peaks and troughs following real-world events (Figure 1) and the distribution of Islamophobia over users is long-tailed (Figure 2) and (b) users exhibit considerable variations in the quality, quantity and temporal patterns of Islamophobic behaviour they engage in. Notably, a small number of users are responsible for most of the Islamophobic hate that we observe. We anticipate that other far right groups, on other platforms, might exhibit slightly different behavioural types and that the prevalence of those types is likely to differ. Nonetheless, we propose that the key argument would hold; the far right is not a single homogeneous group of Islamophobes but a heterogeneous mix of individuals who exhibit systematically different behaviours.

Our findings not only contribute to academic knowledge but also inform policy discussions about how to tackle the spread of divisive and polarising hateful narratives, and the conflict that can ensue due to them. Notably, in a 2019 report, the UK's Commission for Countering Extremism drew an explicit link between online hate and extremist behaviour through their concept and definition of 'hateful extremism' (Commission for Countering Extremism, 2019). Similarly, the 'Online Harms White Paper' from the UK's Department for Digital, Culture, Media & Sport lists online hate as one of the several areas of harm that a regulator should have responsibility for, alongside misinformation and terrorist and extremist content (HM Government, 2019).

The policy implications of our work contribute to, and help advance, these agendas in three main ways. First, the results show opportunities for platforms, such as Twitter, to adopt a more holistic approach to content moderation, which focuses on the behavioural patterns of users rather than just the individual bits of content they produce. Evaluating just whether one bit of content is considered hateful is likely to overlook the broader context in which it is expressed, and the wider array of content that an individual is producing, sharing and engaging with. This is particularly important when making 'edge case' decisions for content or deciding between various available moderation strategies – it may be that quarantines, warning notices, demonetisation and counter speech are more appropriate responses than just bans. Platforms which make longer-term assessments of users, potentially incorporating a long period of data (e.g. 12 months), could make more sophisticated and nuanced assessments, ensuring that a better balance is struck between protecting users' right to free speech and users' right to be protected from harmful hate. Put simply, adopting a user-driven approach to moderation may ultimately be more effective than taking a content-driven approach: targeting the most persistent and overtly hateful users could be used as a 'shortcut', allowing resource-limited moderators to address the primary source of online hate. Since this research was first conducted (2017), platforms have increasingly adopted such an approach, as reflected in the decision by Twitter to ban far right influencers, such as Tommy Robinson in 2018, and Candace Owens and Katie Hopkins in 2020.

Second, the results could be used by policymakers to prioritise resource allocation to counter hate speech and provide support to victims. This would require further research to understand how different types of hate, articulated by different users, drive different responses from victims and wider society – and may, in turn, require different levels of resourcing to be handled. Very different challenges are posed by the infrequent and more nascent Islamophobia expressed by users in the 'Low' type compared with the constant and overt Islamophobia expressed by users in Very High. In particular, the 'Very High' users are potentially more likely to be the purveyors of hate that genuinely 'stir up' intergroup tensions due to the frequency and ferocity of their output. Finally, the Escalating and De-escalating types present opportunities for policymakers to both better understand and tackle far right extremism. Further research would benefit by investigating why these individuals change their behaviour and, in the case of the Escalating type, developing ways of intervening at an early stage.

The policy implications presented here could be evaluated further by investigating more online spaces, such as niche platforms like 8chan, Discord and some communities on Reddit, to see whether similar behavioural patterns are observed and to identify cross-platform patterns of escalation and de-escalation of Islamophobia.

There are several limitations of the current research. The LM model is fit on a simplified representation of the data, comprising just the strongest expression that each user tweets in each time period: the volume of tweets is not modelled directly. Nonetheless, the model performs well at capturing differences in not only the strength but also the volume of Islamophobic tweets. This is because there is an underlying association between the two, which the LM model picks up on as we fit a reasonably large number of time periods. In the future, this could be addressed through using, for instance, a continuous multivariate LM model. Evaluating performance with an unsupervised method is inherently difficult, as many standard evaluative approaches cannot be taken. Our results indicate the existence of several distinct user types, which have been verified by statistical significance testing. Future work should aim to increase the robustness of the modelling and to verify the findings externally.

This research provides a new way of characterising far right actors online and also directly informs policy discussions around how their hateful behaviour can be tackled. Using a mixture of social and computational scientific methodologies, our primary contribution is to identify the internal heterogeneity of the far right and establish the need for more nuanced assessments of hateful far right behaviour online. We have elaborated the policy implications of our analyses, which are increasingly important given the growing threat posed by far right activity.

## Notes

1. We provide the IDs of the 5.2 million tweet data set at: https://zenodo.org/record/3701589#.YDTZfhP7SUo.
2. We repeat all of our analyses on the full data set, without any users removed for high volume tweeting, and find similar results.
3. The average of 89% for 'None' is lower than the value reported in Figure 1(a).

## Disclosure statement

## Funding

This work was supported by Economic and Social Research Council [grant number ES/J500112/1].

## Data availability statement

The data that support the findings of this study are available at https://zenodo.org/record/3701589#.YDTZfhP7SUo.

## ORCID

*Bertie Vidgen*  http://orcid.org/0000-0002-7892-0814
*Taha Yasseri*  http://orcid.org/0000-0002-1800-6094

## References

Allen, C. (2013). Passing the dinner table test: Retrospective and prospective approaches to tackling Islamophobia in Britain. *SAGE Open*, *3*(2), 1–10. doi:10.1177/2158244013484734

Allen, C. (2015). 'People hate you because of the way you dress': Understanding the invisible experiences of veiled British muslim women victims of Islamophobia. *International Review of Victimology*, *21*(3), 287–301. doi:10.1177/0269758015591677

Allen, C. (2017). Islamophobia and the problematization of mosques: A critical exploration of hate crimes and the symbolic function of 'old' and 'new' mosques in the United Kingdom. *Journal of Muslim Minority Affairs*, *37*(3), 294–308. doi:10.1080/13602004.2017.1388477

All Party Parliamentary Group on British Muslims. (2018). *Islamophobia defined: The inquiry into a working definition of Islamophobia*. London: House of Commons.

Awan, I. (2016). Islamophobia on social media: A qualitative analysis of the facebook's walls of hate. *International Journal of Cyber Criminology*, *10*(1), 1–20. doi:10.5281/zenodo.58517

Awan, I., & Zempi, I. (2015). *We fear for our lives : offline and online experiences of anti-Muslim hostility*. London: Tell Mama.

Awan, I., & Zempi, I. (2016). The affinity between online and offline anti-Muslim hate crime: Dynamics and impacts. *Aggression and Violent Behaviour*, *27*(1), 1–8. doi:10.1016/j.avb.2016.02.001

Awan, I., & Zempi, I. (2017). 'I will blow your face off' - virtual and physical world Anti-Muslim hate crime. *British Journal of Criminology*, *57*(2), 362–380. doi:10.1093/bjc/azv122

Baele, S., Brace, L., & Coan, T. (2020). Uncovering the far-right online ecoystem: An analytical framework and research agenda. *Studies in Conflict & Terrorism*. doi:10.1080/1057610X.2020.1862895

Bakali, N. (2019). *The far right's love affair with Islamophobia: Vol. 1. Yaqeen*: Institute for Islamic Research.

Bartolucci, F., Farcomeni, A., & Pennoni, F. (2010). "An Overview of Latent Markov Models for Longitudinal Categorical Data." *ArXiv:1003*.2804, 1–36.

Benford, R., & Snow, D. (2000). Framing processes and social movements: An overview and assessment. *Annual Review of Sociology*, *26*(1), 611–639.

Beydoun, K. (2016). Islamophobia: Toward a legal definition and framework. *Columbia Law Review*, *116*(1), 108–125.

Biggs, M., & Knauss, S. (2012). Explaining membership in the British National Party: A multilevel analysis of contact and threat. *European Sociological Review*, *28*(5), 633–646. doi:10.1093/esr/jcr031

Bleich, E. (2011). What is islamophobia and how much is there? Theorizing and measuring an emerging comparative concept. *American Behavioral Scientist*, *55*(12), 1581–1600. doi:10.1177/0002764211409387

Bleich, E. (2012). Defining and researching Islamophobia. *Review of Middle East Studies*, *46*(2), 180–189.

Bliuc, A. M., Faulkner, N., Jakubowicz, A., & McGarty, C. (2018). Online networks of racial hate: A systematic review of 10 years of research on cyber-racism. *Computers in Human Behavior*, *87*(1), 75–86. doi:10.1016/j.chb.2018.05.026

Brown, A. (2018). What is so special about online (as compared to offline) hate speech? *Ethnicities*, *18*(3), 297–326. doi:10.1177/1468796817709846

Burnap, P., Williams, M., Sloan, L., Rana, O., Housley, W., Edwards, A., … Voss, A. (2014). Tweeting the terror: Modelling the social media reaction to the Woolwich terrorist attack. *Social Network Analysis and Mining*, *4*(1), 1–14.

Chakraborti, N., & Zempi, I. (2012). The veil under attack: Gendered dimensions of islamophobic victimization. *International Journal of Victimology*, *18*(3), 269–284. doi:10.1177/0269758012446983

Chandrasekharan, E., Pavalanathan, U., Gilbert, E., Srinivasan, A., Glynn, A., & Eisenstein, J. (2017). You can't stay here. *Proceedings of the ACM on Human-Computer Interaction*, *1*(2), 1–22. doi:10.1145/3134666

Chatzakou, D., Kourtellis, N., Blackburn, J., De Cristofaro, E., Stringhini, G., & Vakali, A. (2017). Hate is not binary: Studying abusive behavior of #GamerGate on Twitter." Retrieved March 11, 2019 from https://arxiv.org/pdf/1705.03345.pdf

Commission for Countering Extremism. (2019). *Challenging hateful extremism*. London: Commission for Countering Extremism.

Copsey, N. (2007). Changing course or changing clothes? Reflections on the ideological evolution of the British National Party 1999-2006. *Patterns of Prejudice*, *41*(1), 61–82. doi:10.1080/00313220601118777

Cutts, D., Ford, R., & Goodwin, M. (2011). Anti-immigrant, politically disaffected or still racist after all? Examining the attitudinal drivers of extreme right support in Britain in the 2009 European elections. *European Journal of Political Research*, *50*(3), 418–440. doi:10.1111/j.1475-6765.2010.01936.x

Davidson, T., Warmsley, D., Macy, M., & Weber, I. (2017). *Automated hate speech detection and the problem of Offensive language*. Proceedings of the 11th International Conference on Web and Social media (pp. 1–4).

Demos. (2017). *Anti-Islamic hate on twitter*. London: DEMOS.

Eatwell, R., & Goodwin, M. (2010). *The new extremism in the 21st century*. Abingdon: Routledge.

Ellis, A., Burchett, W., Harrar, S., & Bathke, A. (2017). Nonparametric inference for multivariate data: The R Package Npmv. *Journal of Statistical Software*, *76*(4), 1–18. doi:10.18637/jss.v076.i04

Evolvi, G. (2018). Hate in a tweet: Exploring internet-based islamophobic discourses. *Religions*, *9*(10), 307. doi:10.3390/rel9100307

Falkheimer, J., & Olsson, E. K. (2015). Depoliticizing terror: The news framing of the terrorist attacks in Norway, 22 July 2011. *Media, War and Conflict*, *8*(1), 70–85. doi:10.1177/1750635214531109

Ford, R., & Goodwin, M. J. (2010). Angry white men: Individual and contextual predictors of support for the British National party. *Political Studies*, *58*(1), 1–25. doi:10.1111/j.1467-9248.2009.00829.x

Froio, C. (2018). Race, religion, or culture? Framing Islam between racism and Neo-racism in the online network of the French far right. *Perspectives on Politics*, *16*(3), 696–709. doi:10.1017/S1537592718001573

Ganesh, B. (2018). The ungovernability of digital hate culture. *Journal of International Affairs*, *71*(2), 30–49.

Golder, M. (2016). Far right parties in Europe. *Annual Review of Sociology*, *19*(1), 477–497. doi:10.1146/annurev-polisci-042814-012441

Goodwin, M. (2007). The extreme right in Britian: Still an 'ugly duckling' but for how long? *The Political Quarterly*, *78*(2), 241–250. doi:10.1111/j.1467-923X.2007.00851.x

Goodwin, M. (2013). Forever a false Dawn? Explaining the electoral collapse of the British National Party (BNP). *Parliamentary Affairs*, *67*(4), 1–20. doi:10.1093/pa/gss062

Goodwin, M., Ford, R., & Cutts, D. (2012). Extreme right foot soldiers, legacy effects and deprivation: A contextual analysis of the leaked British National Party (BNP) membership list. *Party Politics*, *19*(6), 887–906. doi:10.1177/1354068811436034

Gottlieb, J. V. (2017). Women and British fascism revisited: Gender, the Far-right, and resistance. *Journal of Women's History*, *16*(3), 108–123.

Hine, G. E., Onaolapo, J., De Cristofaro, E., Kourtellis, N., Leontiadis, I., Samaras, R., … Blackburn, J. (2017). *Kek, cucks, and God emperor Trump: A measurement study of 4chan's Politically incorrect forum and its effects on the web*. Proceedings of the 11th International Conference on Web and Social media (pp. 92–101).

HM Government. (2019). *Online Harms White paper*. London: Department of Digital, Culture, Media and Society.
Home Office. (2012). *Challenge it, report it, stop it: The government 's plan to tackle hate crime*. London: Home Office.
Home Office. (2016). *Action against hate: The UK Government's plan for tackling hate crime*. London: Home Office.
Home Office. (2018). *Action against hate: The UK Government's plan for tackling hate crime – two years On*. London: Home Office.
Home Office. (2019). Hate crime, England and Wales, 2018 to 2019. *Home Office Statistical Bulletin*.
Hope Not Hate. (2017). *State of hate*. London: Hope Not Hate.
Ignazi, P. (2003). *Extreme right parties in Western Europe*. Oxford: Oxford University Press. https://doi.org/10.1093/0198293259.001.0001
Ingham-Barrow, I. (2018). In I. Ingham-Barrow (Ed.), *More than words: Approaching a definition of Islamophobia*. London: MEND.
John, P., & Margetts, H. (2009). The latent support for the extreme right in British politics. *West European Politics*, 32(3), 496–513. doi:10.1080/01402380902779063
Klug, B. (2012). Islamophobia: A concept comes of age. *Ethnicities*, 12(5), 665–681. doi:10.1177/1468796812450363
Larsson, A., & Hallvard, M. (2015). Bots or journalists? News sharing on Twitter. *Communications*, 40(3), 361–370. doi:10.1515/commun-2015-0014
Leader Maynard, J., & Benesch, S. (2016). Dangerous speech and dangerous ideology: An integrated model for monitoring and prevention. *Genocide Studies and Prevention*, 9(3), 70–95. doi:10.5038/1911-9933.9.3.1317
Macklin, G. (2013). Transnational networking on the far right: The case of Britain and Germany. *Western European Politics,* 36(1), 176–198.
Margetts, H., John, P., Hale, S., & Yasseri, T. (2015). *Political turbulence: How social media shape collective action*. Oxford: Princeton University Press.
Marquez, J. J., Downey, A., & Clement, R. (2015). Walking a mile in the user's shoes: Customer journey mapping as a method to understanding the user experience. *Internet Reference Services Quarterly*, 20(3–4), 135–150. doi:10.1080/10875301.2015.1107000
Marranci, G. (2006). Multiculturalism, Islam and the clash of civilisations theory: Rethinking islamophobia. *Culture and Religion*, 5(1), 105–117. doi:10.1080/0143830042000200373
Matsuda, M., Lawrence, C., Delgado, R., & Crenshaw, K. (1993). *Words that wound: Critical race theory, assaultive speech and the first amendment*. New York: Routledge.
McCauley, C., & Moskalenko, S. (2008). Mechanisms of political radicalization: Pathways toward terrorism. *Terrorism and Political Violence*, 20(3), 415–433. doi:10.1080/09546550802073367
Mohideen, H., & Mohideen, S. (2008). The language of Islamophobia in Internet articles. *Intellectual Discourse*, 16(1), 73–87. http://www.iium.edu.my/intdiscourse/index.php/islam/article/viewArticle/31
Morrow, E., & Meadowcroft, J. (2019). The rise and fall of the English Defence League: Self-governance, marginal members and the far right. *Political Studies*, 67(3), 539–556.
Mudde, C. (2014). Fighting the system? Populist radical right parties and party system change. *Party Politics*, 20(2), 217–226. doi:10.1177/1354068813519968
Mudde, C. (2009). Populist radical right parties in Europe redux. *Political Studies Review,* 7(3), 330–337.
Rhodes, J. (2011). 'It's not just them, It's whites as well': Whiteness, class and BNP support. *Sociology*, 45(1), 102–117. doi:10.1177/0038038510387191
Richardson, J., & Wodak, R. (2017). Recontextualising fascist ideologies of the past: Right-Wing discourses on employment and nativism in Austria and the United Kingdom. *Critical Discourse Studies*, 6(4), 251–267. doi:10.1080/17405900903180996
Runnymede Trust. (2017). *Islamophobia: Still a challenge for Us All*. London: Runnymede Trust.
Rydgren, J. (2007). The sociology of the radical right. *Annual Review of Sociology*, 33(1), 241–262. doi:10.1146/annurev.soc.33.040406.131752

Rydgren, J. (2008). Immigration sceptics, xenophobes or racists? Radical right-Wing voting in six west European countries. *European Journal of Political Research*, *47*(6), 737–765. doi:10.1111/j.1475-6765.2008.00784.x

Rydgren, J. (2010). Sweden: The Scandinavian exception. In D. Albertazzi and D. McDonnell (Eds.), *Twenty-first century populism* (pp. 135–150). London: Palgrave Macmillan.

Sayyid, S. (2014). A measure of Islamophobia. *Islamophobia Studies Journal*, *2*(1), 10–25. doi:10.13169/islastudj.2.1.0010

Severs, G. J. (2017). The 'obnoxious mobilised minority': Homophobia and homohysteria in the British National Party, 1982–1999. *Gender and Education*, *29*(2), 165–181. doi:10.1080/09540253.2016.1274384

Shepherd, T., Harvey, A., Jordan, T., Srauy, S., & Miltner, K. (2015). Histories of hating. *Social Media & Society*, *1*(2), 1–10. doi:10.1177/2056305115603997

Sheridan, L. (2006). Islamophobia pre– and post–September 11th, 2001. *Journal of Interpersonal Violence*, *21*(3), 317–336. doi:10.1177/0886260505282885

Simpson, R. M. (2013). Dignity, harm, and hate speech. *Law and Philosophy*, *32*(6), 701–728. doi:10.1007/s10982-012-9164-z

Spedicato, G. A., & Signorelli, M. (2013). "The Markovchain Package: A Package for Easily Handling Discrete Markov Chains in R." *R CRAN*, 1–67. ftp://sunsite2.icm.edu.pl/site/cran/web/packages/markovchain/vignettes/an_introduction_to_markovchain_package.pdf.

Tell MAMA. (2017). *A constructed threat: Identity, prejudice and the impact of Anti-Muslim hatred*. London: Tell MAMA.

Trilling, D. (2012). *Bloody nasty people: The rise of Britain's far right*. London: Verso.

van Rijsbergen, C. J. (1979). *Information retrieval*. London: Butterworths.

Varol, O., Ferrara, E., Davis, C. A., Menczer, F., & Flammini, A. (2017). *Online human-bot interactions: Detection, estimation, and characterization*. Proceedings of the 11th International Conference on web and social media (pp. 280–290).

Veugelers, J., & Magnan, A. (2005). Conditions of far-right strength in contemporary Western Europe: An application of Kitschelt's theory. *European Journal of Political Research, 44*(6), 837–860.

Vidgen, B., Tromble, R., Harris, A., Hale, S., Nguyen, D., & Margetts, H. (2019). *Challenges and frontiers in abusive content detection*. Proceedings of the 3rd workshop on Abusive language online (ACL).

Vidgen, B., & Yasseri, T. (2020). Detecting weak and strong Islamophobic hate speech on social media. *Journal of Information Technology & Politics*, *17*(1), 66–78. doi:10.1080/19331681.2019.1702607

Watanabe, H., Bouazizi, M., & Ohtsuki, T. (2018). Hate speech on twitter: A pragmatic approach to collect hateful and offensive expressions and perform hate speech detection. *IEEE Access*, *6*, 13825–13835. doi:10.1109/ACCESS.2018.2806394

Williams, M., & Burnap, P. (2016). Cyberhate on social media in the aftermath of woolwich: A case study in computational criminology and big data. *British Journal of Criminology*, *56*(1), 211–238. doi:10.1093/bjc/azv059

Zannettou, S., Caulfield, T., De Cristofaro, E., Kourtellis, N., Leontiadis, I., Sirivianos, M., … Blackburn, J. (2017). *The web centipede: Understanding how web communities influence each other through the lens of mainstream and alternative news sources*. Proceedings of the ACM Internet Measurement conference (pp. 1–14). https://doi.org/10.1145/3131365.3131390

Zúquete, J. P. (2008). The European extreme-right and Islam: new directions? *Journal of Political Ideologies*, *13*(3), 321–344. doi:10.1080/13569310802377019

## Appendix

### *1. Islamophobia classifier*

The classifier for Islamophobic content is described in detail in (Vidgen & Yasseri, 2020). It was trained on a newly annotated data set of 4000 tweets and achieves balanced accuracy of 0.83 and a micro-F1 score of 0.78 when tested on an unseen 300 tweet data set. Precision is 0.78, which is far above the 0.7 minimum recommended by van Rijsbergen for empirical research (van Rijsbergen, 1979). The main sources of classification error are (1) confusing implicit and explicit Islamophobia and (2) confusing non-Islamophobic with implicit Islamophobic, specifically content which (a) discusses Muslims and Islam but in a non-hateful way and (b) is hateful against another target, such as immigrants or minority ethnic groups. Overall, the classifiers' performance compares well with other ternary multi-class classifiers for abusive content and is suitable for empirical research (Davidson, Warmsley, Macy, & Weber, 2017; Watanabe, Bouazizi, & Ohtsuki, 2018).

## 2. Latent Markov model fitting

LM models are tested for 1–12 latent states and evaluated with AIC and BIC. Both measures are closely aligned and indicate that a range of between 3 and 7 latent states is optimal, as shown in Figure A1. Fitting a number of latent states towards the top-end of the indicated range (e.g. 5–7) is problematic because each latent state accounts for a specific range of behaviours. Because the states are highly tailored to small subsets of users, it is less likely that users will transition between states. Therefore, the transitional probabilities become very high for remaining in the same state. For instance, in a model with five latent states, the transitional probabilities for remaining in the same state are all over 0.98.

## 3. Number of clusters: k-modes fitting

From the output of the LM model, we test for between 2 and 20 clusters with the *k*-modes algorithm (referred to as 'user types' in the main body of the paper), evaluated by measuring the Within Sum of Squares. The results indicate a range of between 5 and 8 clusters is optimal. This is shown in Figure A2.

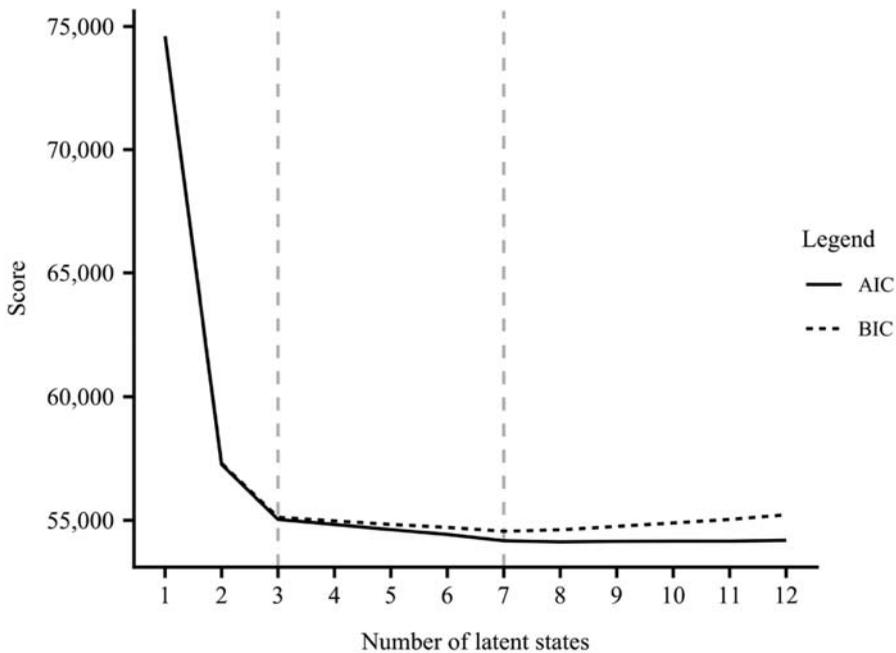

**Figure A1.** Results of fitting the optimal number of latent states in the LM model, using AIC and BIC.

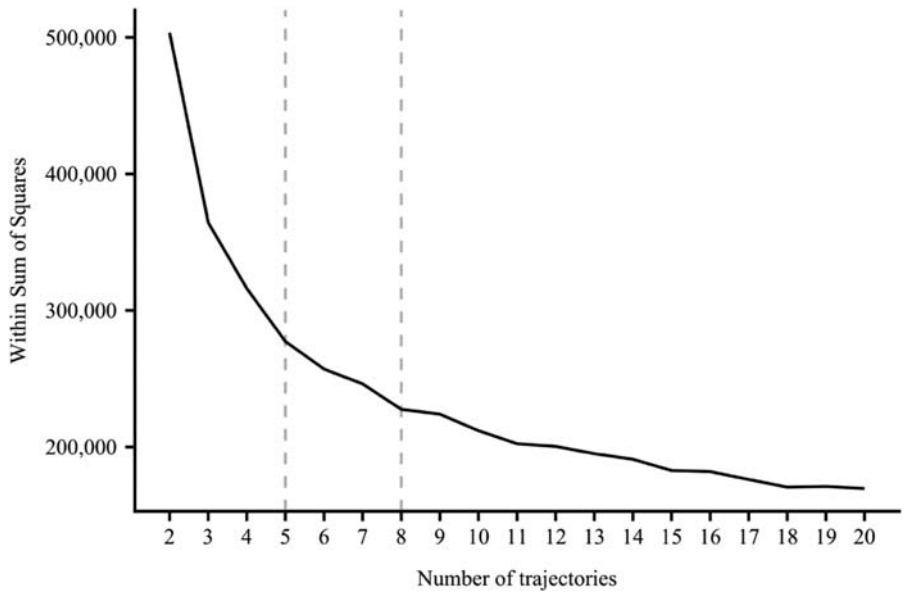

**Figure A2.** Results of fitting for the optimal number of user types, using Within Sum of Squares.

## 4. Quantitative differences between the seven user types

**Table A1.** Quantitative differences in None, Implicit and Explicit Islamophobic tweeting between the seven types of Islamophobia.

| User type | None | | Implicit | | Explicit | | All |
|---|---|---|---|---|---|---|---|
| | Mean (None) | (None) Standard deviation | Mean (Implicit) | Standard deviation (Implicit) | Mean (Explicit) | Standard deviation (Explicit) | Total number of tweets |
| None | 28 (100%) | 139 | 0 (0%) | 0 | 0 (0%) | 0 | 28 |
| Very low | 128 (97%) | 266 | 3 (3%) | 7 | 0 (0%) | 0 | 131 |
| Low | 263 (94%) | 541 | 12 (4%) | 32 | 6 (2%) | 13 | 281 |
| High | 2,137 (89%) | 2,240 | 182 (8%) | 223 | 70 (3%) | 119 | 2,389 |
| Very High | 4,762 (79%) | 3,084 | 844 (14%) | 692 | 439 (7%) | 386 | 6,045 |
| Escalating | 1,140 (90%) | 1,335 | 92 (7%) | 166 | 37 (3%) | 77 | 1,269 |
| De-escalating | 1,211 (83%) | 1,668 | 160 (11%) | 271 | 83 (6%) | 144 | 1,454 |
| Cohort average | 854 (89%)[3] | 1,882 | 110 (7%) | 330 | 54 (3%) | 176 | 1,018 |